# Width-Dependent Band Gap in Armchair Graphene Nanoribbons Reveals Fermi Level Pinning on Au(111).

*Néstor Merino-Díez,* [‡,†] *Aran Garcia-Lekue,* [‡,§]*, Eduard Carbonell-Sanromà,* [†] *Jingcheng Li,* [⊥,†] *Martina Corso,* [†,§,⊥] *Luciano Colazzo,* [∇] *Francesco Sedona,* [∇] *Daniel Sánchez-Portal,* [‡,⊥] *Jose I. Pascual,* [†,§] *and Dimas G. de Oteyza* [‡,§,⊥].

[‡]Donostia International Physics Center (DIPC), 20018 Donostia – San Sebastián, Spain

[†]CIC nanoGUNE, Nanoscience Cooperative Research Center, 20018 Donostia – San Sebastián, Spain

[§] Ikerbasque, Basque Foundation for Science, 48013 Bilbao, Spain

[⊥] Centro de Física de Materiales (CSIC/UPV-EHU) - Materials Physics Center, 20018 Donostia – San Sebastián, Spain

[∇] Dipartimento di Scienze Chimiche, Università di Padova, 35131 Padova, Italy

Author(s) address: d_g_oteyza@ehu.es

ABSTRACT

We report on the energy level alignment evolution of valence and conduction bands of armchair-oriented graphene nanoribbons (aGNR) as their band gap shrinks with increasing width. We use 4,4''-dibromo-*para*-terphenyl as molecular precursor on Au(111) to form extended poly-*para*-phenylene nanowires, which can be fused sideways to form atomically precise aGNRs of varying widths. We measure the frontier bands by means of scanning tunneling spectroscopy, corroborating that the nanoribbon's band gap is inversely proportional to their width. Interestingly, valence bands are found to show Fermi level pinning as the band gap decreases below a threshold value around 1.7 eV. Such behavior is of critical importance to understand the properties of potential contacts in graphene nanoribbon-based devices. Our measurements further reveal a particularly interesting system for studying Fermi level pinning by modifying an adsorbate´s band gap while maintaining an almost unchanged interface chemistry defined by substrate and adsorbate.

Graphene nanoribbons (GNRs) have long been proposed as extremely interesting materials for a variety of applications, ranging from their key role in composites with mechanically robust films and high gas barrier efficiencies,[1] to thermoelectric devices,[2,3] capacitors,[4,5] batteries,[6] photodetectors,[7] transistors [8] or directly integrated circuits.[9] Among the most appealing attributes that make GNRs so interesting, we find the greatly tunable properties they display as a function of their precise atomic structure. While this holds enormous interest for many of the applications mentioned above, it also underlines the need for atomic precision in their synthesis if the full potential of GNRs is to be exploited. That was achieved for the first time in 2010 with the bottom-up synthesis of armchair oriented nanoribbons (aGNRs) with 7 dimer lines across their width (7-aGNR).[10] Ever since, following a similar "on-surface synthesis" approach, great efforts are being devoted to the synthesis of GNRs with different widths,[11–13] edge orientations,[14,15] dopants,[16–20] as well as heterostructures thereof.[21–24]

The increasing pool of available GNRs with well-defined structures has allowed the subsequent characterization of their fundamental electronic properties, as well as the correlation with their performance when integrated into devices like field effect transistors.[25,26] Interestingly, it has been found that the device performance is strongly dominated by contact effects, in particular by the Schottky barrier at the GNR-contact interface. As opposed to studies of other GNR properties like their band gaps, systematic studies of the energy level alignment between GNRs and common contact materials are still missing in spite of their determining role in the ultimate response of GNR-based devices. In this work we amend our understanding of such interface energetics between GNRs and gold, in particular Au(111) surfaces.

Armchair oriented nanoribbons are known to display a width-dependent band gap. Calculations reveal that they can be classified into three different subfamilies depending on the number of dimer lines $p$ across their width ( $3p$, $3p+1$, or $3p+2$), their band gaps being inversely proportional to the width within each of those families.[27–29] The reported band gap values of the various aGNRs synthesized to date confirm this picture with scattered points along the predictions for each of the GNR families.[29,30] In this work, we provide a systematic study of the band gap and energy level alignment of GNRs focused on the $3p$ family, addressing from the smallest possible GNR (3-aGNR) to its four immediately following sister-structures (6-, 9-, 12-, and 15-aGNRs). Starting from the synthesis of poly-*para*-phenylene wires (PPP or 3-aGNR) on Au(111), subsequent annealing drives their lateral fusion and results in the required atomically precise GNRs of varying width.[31] The following characterization by scanning tunneling spectroscopy (STS) and density functional theory (DFT) calculations reveals, in addition to the width-dependent band gap, the onset of Fermi level pinning for widths ≥ 6 dimer lines.

**RESULTS AND DISCUSSION**

We use 4,4''-dibromo-*p*-terphenyl (DBTP) as molecular precursor to synthesize the structures described in this work. Figure 1 summarizes the stepwise synthesis of aGNRs starting from the precursor. A submonolayer of DBTP is initially deposited on a Au(111) single crystal held at room temperature. After post-deposition annealing above 250°C, the precursor undergoes Ullmann-like coupling,[31,32] yielding poly-*para*-phenylene (PPP) nanowires. Figure 1b shows a scanning tunneling microscopy (STM) image of a representative sample. As previously reported,[31] these nanowires are highly aligned, separated by arrays of bromine atoms in between them (see Figure S1a), and present impressive lengths of up to 200 nm. PPP has been occasionally termed as a 3-aGNR and would thus fit into a family of $3p$-aGNRs. However, as previously concluded from near edge X-ray absorption fine structure (NEXAFS) measurements, it displays a non-planar structure on Au.[20] While not recognizable in constant current ("topographic") STM imaging (Figure 1b), mapping of the density of states (DOS) at the onset of valence and conduction bands clearly reveals a modulated intensity that mirrors an alternating tilt of subsequent phenyl rings (Figure S1). The non-planarity of gold-supported PPP is also confirmed by DFT calculations even when using functionals that incorporate the vdW interaction, although reduced with respect to that in gas phase PPP. As discussed later, this non-planar structure leads to a substantial difference in the electronic properties of the polymer with respect to the rest of wider (and planar) 3p-aGNRs.

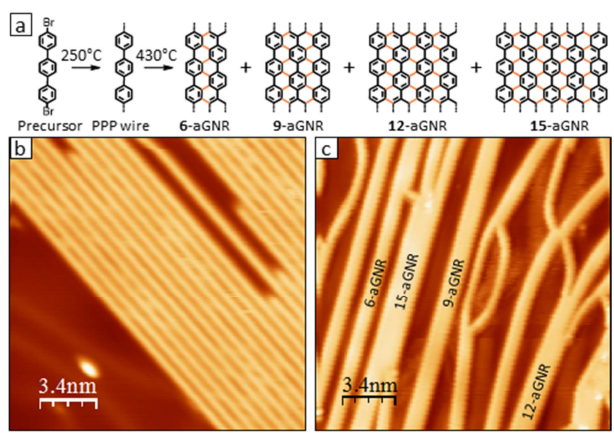

**Figure 1. (a)** Schematic synthesis from DBTP molecular precursor on Au(111). STM topography images of **(b)** PPP nanowires (24.6 nm x 24,6 nm, It = 0.22 nA, Vs = 1.0 V) and **(c)** a-GNRs (24.6 nm x 24,6 nm, It = 0.22 nA, Vs = -1.7 V) where text inlets indicate the different widths of aGNRs.

Above 430°C, PPP nanowires merge sideways through dehydrogenation,[31] forming aGNRs of different width as shown in Figure 1c. This synthetic step relies on random diffusion effects imposed by the high temperature and results in an interconnected network of nanoribbons of different widths. There is a significant number of curved aGNRs, but most of them preserve the preceding nanowire´s straight orientation. The final nanoribbon width depends on the number of nanowires getting fused. Discretized by the intrinsic width of PPP nanowires (3-aGNR), we find 6-, 9-, 12- and 15-aGNR are formed from the fusion of two, three, four and five nanowires, respectively. Having them all next to each other on the same Au(111) surface reveals itself as an excellent testbed to measure and compare their respective electronic properties.

The aGNR´s electronic properties have been characterized by STS, measuring both differential conductance (dI/dV) point spectra and dI/dV maps at various sample bias voltages (Figure 2). The point spectra were taken over the sides (where the measured GNR signal is highest[22,33,34]) of straight GNR segments, featuring a well-defined width and defect free structure over substantial lengths. While segments of 6- and 9-aGNRs easily exceed tens of nanometers, the wider the ribbons the shorter the average segments are. To avoid notable band gap variations from finite length effects, spectra are only taken into account from segments longer than 4 nm, *i.e.* ~9 unit cells. While the convergence behavior of GNR band gap with its width depends on a series of aspects like e.g the ribbon´s own polarizability and thus on its particular structure, for the GNRs studied here the probed electronic properties at these lengths can be considered to be close to those of an infinite ribbon.[35,36] Figure 2a displays representative STS spectra for ribbons with different widths, together with background spectra of the bare Au(111) surface nearby, as a reference.

All STS spectra show a clear conductance rise at positive bias, which is attributed to the onset of the conduction band (CB). The position of these bands is observed at different energies as a function of the ribbons' width: the wider the ribbon, the closer to $E_F$ its conduction band is. In this way, the onsets range from 0.86 eV, for the wider 15-aGNR, to 1.47 eV for the narrower 6-aGNR. Regarding the filled states ($V_S < 0$), the first nanoribbon-related feature in the spectra is detected at bias values close to -0.2 eV for all measured widths. We associate it to the corresponding onset of the aGNR valence band (VB). A more detailed analysis reveals that the valence band onset follows a similar, although less pronounced, width-related trend as the one observed in the conduction band case. The valence bands are slightly closer to $E_F$ as the GNRs become wider, ranging from -0.17 eV to -0.23 eV. It is worth noting that the detection of the VB onset is particularly difficult on Au(111) since the surface state signal overlaps with the 3p-AGNRs spectroscopic features. Additionally, the density of states of the valence band of 3p-AGNRs decays particularly fast along the direction perpendicular to the GNR plane.[12,34] Both of these reasons severely complicate its detection by STS. Nevertheless, our analysis of both valence and conduction bands is in excellent agreement with the recently reported electronic structure and energy level alignment of 9-aGNRs synthesized selectively on Au(111) from appropriate precursors.[12]

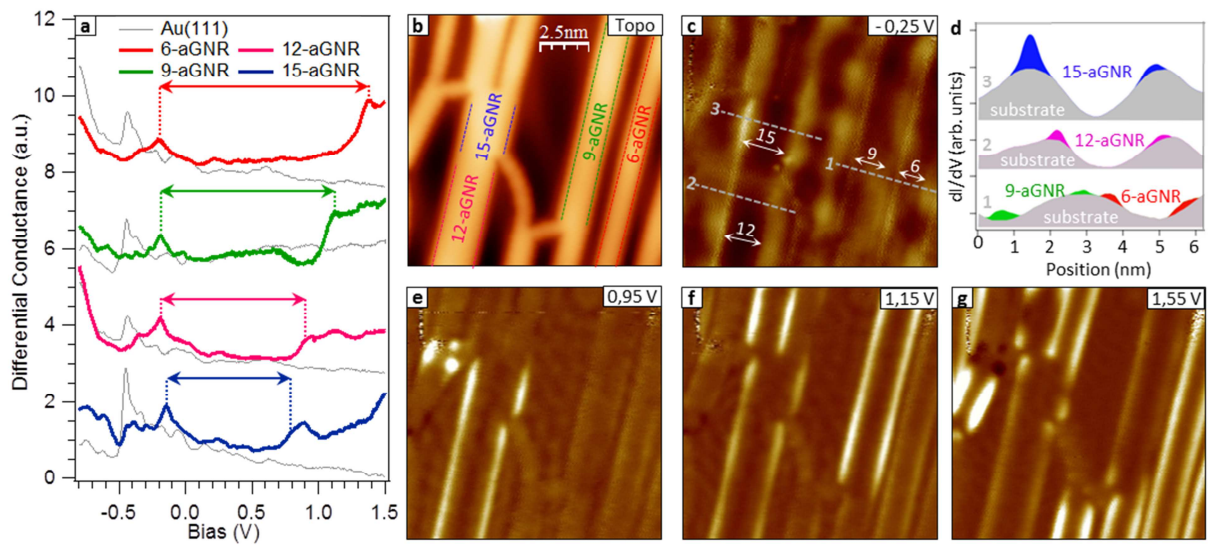

**Figure 2. (a)** Spectra recorded on 6-aGNR (red), 9-aGNR (green), 12-aGNR (pink) and 15-aGNR (blue) where Au(111) signal (black) is added to every spectrum as background reference (open-feedback parameters: $V_s$ = 1.55 V, $I_t$ = 1.4 - 10 nA; modulation voltage $V_{rms}$ = 10 mV). **(b)** STM topography image (12,4 nm x 12.4 nm; $V_s$ = -1.1 V; $I_t$ = 0.61 nA). **(c)** Conductance map near the valence band onset ($V_s$ = -0.25 V), with white arrows as a guide to the eye

to the intensity along each aGNR edge. **(d)** Profiles across the conductance map in (c), highlighting the contribution from the VB of differently wide GNRs on top of the dominating Au(111) surface state contribution. Conductance maps near the conduction band onsets for **(e)** 15-aGNRs and 12-aGNRs ($V_s$ = 0.95 V) **(f)** 9-aGNR ($V_s$ = 1.15 V) and **(g)** 6-aGNR ($V_s$ =1.55 V). The displayed spectra have been taken from ribbons on different sample locations (not necessarily on those of panels b-g). Differences in the reference spectra relate to different tips and positions. The comparable intensity in 12- and 15-aGNRs' edges in image (d) probably relates to the limited length of the 15-aGNR segment. Its reduced length causes an increased band gap and thus leads to an upshift in the energy of the conduction band onset, making it overlap with that of the longer 12-aGNR. Size and setpoint for all conductance images were 12,4 nm x 12.4 nm and $I_t$ = 0.61 nA, respectively.

The conclusions derived above from our dI/dV point spectra are supported by dI/dV maps obtained at different biases (Figure 2c-g). When probing the empty states ($V_S > 0$), as one increases the energy, only the oscillations associated with scattered surface state electrons are initially visible (Fig. S2). However, there is an energy threshold at which an increased conductance starts becoming visible along the sides of the ribbons, first for the wider ribbons (Fig. 2e) and as the energy increases also for 9-aGNRs (Fig. 2f) and 6-aGNRs (Fig. 2g). Such intensity in conductance maps is related to the local density of states (LDOS) of aGNR bands,[11,34] and the threshold energies at which the increased conductance sets in, is in agreement with the width-dependent band onsets determined from the point spectra. In the case of the filled states ($V_S < 0$), conductance maps at -0.25 V reveal the strong intensity of the scattered surface state. Superimposed to it, we observe a weak but recognizable intensity distributed along the edges of every nanoribbon (Fig. 2c) further highlighted in the profiles displayed in Fig. 2d and Fig. S2. This agrees with the VB onsets being at similar energies for all GNRs, as observed in the spectra (Figure 2a). Besides the frontier bands, our spectroscopy measurements reveal an additional nanoribbon-related signal at higher negative bias (Fig. S3). We associate it to the second valence band (VB-1), which is easier to be probed by the tip of the STM than the VB. Its slower decay away from the nanoribbon plane explains,[12,34] together with the absence of the Au(111) surface state at these energies, its stronger and more easily measurable signal (Figure S3). For a clear visualization of the overall data, table 1 summarizes the energies of all band onsets averaged over several measurements on different aGNRs, as well as the associated band gaps.

**Table 1.** Average spectroscopic results on VB-1, VB, CB and resulting bandgaps.

| Structure | VB-1 (eV) | VB (eV) | CB (eV) | Bandgap (eV) |
|---|---|---|---|---|
| **PPP wire** |  | -1.09 ± 0.05 | 2.14 ± 0.06 | 3.23 ± 0.08 |
| **6-aGNR** | -1.73 ± 0.04 | -0.23 ± 0.08 | 1.47 ± 0.05 | 1.69 ± 0.10 |
| **9-aGNR** | -1.17 ± 0.06 | -0.20 ± 0.05 | 1.14 ± 0.05 | 1.35 ± 0.07 |
| **12-aGNR** | -0.84 ± 0.04 | -0.18 ± 0.04 | 0.96 ± 0.04 | 1.13 ± 0.05 |
| **15-aGNR** | -0.66 ± 0.09 | -0.17 ± 0.03 | 0.86 ± 0.03 | 1.03 ± 0.04 |

The average band gaps obtained for the differently wide aGNRs are displayed in Figure 3, where PPP nanowires (analyzed in Figure S1) are included for comparison as 3-aGNRs. The values fit into the predicted range for gold-supported nanoribbons of similar widths[29] and their smooth evolution is also in line with predictions.[27,29] However, 3-aGNRs clearly stick out of the smoothly varying trend of a monotonously decreasing band gap with incremental width within the 3p-aGNR family. The reason behind this is that, in addition to the larger band gap associated with its narrower width, it is the only non-planar structure. As a result, the degree of conjugation is reduced,[37] causing an anomalous band gap increase for this particular structure different from the wider graphene nanoribbons. Figure 3 also displays the average onset energies of valence and conduction band for the differently wide aGNRs. It becomes immediately clear that GNRs display an overall p-type alignment on Au(111) and that for 6- and wider aGNRs, as the bandgap decreases, the conductance band onset approaches the Fermi level much faster than the valence band.

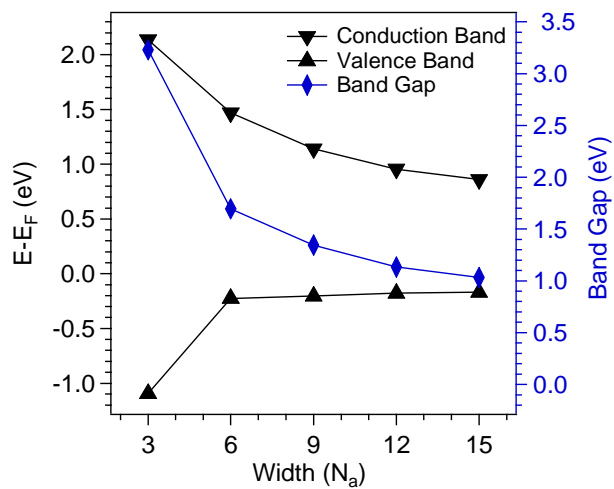

**Figure 3.** Average spectroscopic results for valence and conduction bands (black lines, left scale) and the resulting band gaps (blue line, right scale).

Both these effects are indeed reproduced by DFT calculations of Au(111)-supported graphene nanoribbons from 3 (PPP wires) to 12 dimer lines. Figure 4 displays the band structure of the various systems, where the diameter of blue symbols is proportional to the weight of each state on the carbon atoms and thus on the GNRs. Next to the band structures, the figure displays the density of states projected onto C. The ribbons show a clear p-type alignment with the Au(111) substrate. In line with our experiment-based observations, after a notable upshift of the valence band from 3-aGNR to 6-aGNR, for wider ribbons the band onset appears close to the Fermi energy at a position fairly independent on the ribbon width. It shifts by only ~ 0.16 eV when going from 6-aGNRs to 12-aGNRs, to be compared with an about three times larger shift of the conduction band for the same widths.

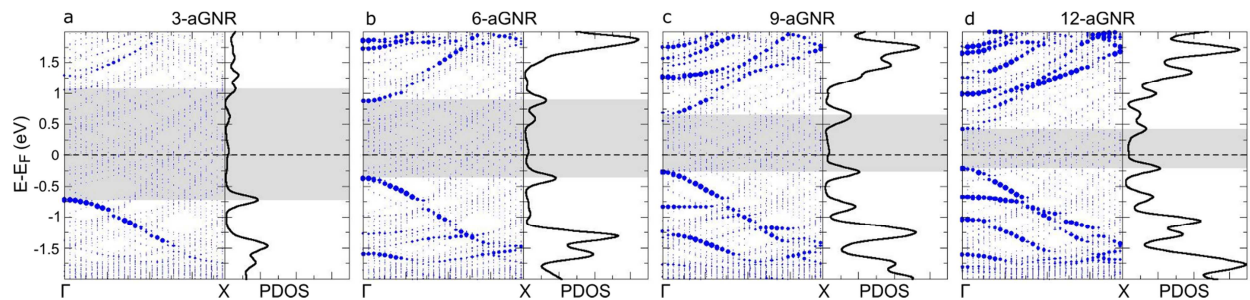

**Figure 4.** Calculated band structure (left) and PDOS (right) for **(a)** 3-aGNRs, **(b)** 6-aGNRs, **(c)** 9-aGNRs and **(d)** 12-aGNRs. The diameter of the blue circles denotes the density of states projected onto the GNR´s carbon atoms. The shadowed areas indicate the respective band gaps.

Experimentally, the p-type alignment of GNRs seems reasonable due to the p-type doping observed for gold-supported graphene,[38–42] which stems from the much larger workfunction of gold as compared to that of graphene. However, calculations have shown the understanding of this alignment not to be trivial, since it depends on the dipole layer formed at the metal-graphene interface and displays a marked dependence on the details of the graphene-substrate interaction and the associated adsorption distance.[43–45] GNRs have a weak interaction with the Au(111) surface, dominated by van der Waals forces. As a weakly interacting semiconducting adsorbate approaches a metal surface, the substrate´s intrinsic surface dipole is modified by an amount $\Delta$ as a result of a variety of processes that include, among others, the surface´s electron cloud redistribution arising from Pauli repulsion with the adsorbate (commonly termed as "pillow effect"), or intrinsic adsorbate´s dipole moments.[46] In the case of aGNRs on Au(111) only the former process is relevant, giving rise to an effective reduction of the substrate workfunction.

Indeed, for extended graphene on Au(111) the effect is such that, although at the equilibrium adsorption height (~3.3 Å) p-doping is predicted by first-principles calculations, for only slightly smaller distances (< 3.2 Å) n-doping is obtained instead.[43,44] Therefore, the strong p-type alignment observed here for the aGNRs is not necessarily obvious. Particularly intriguing is the observation, in agreement with experiment, of very asymmetric shifts of the VB and CB onsets in the DFT calculations. Although the VB approaches the Fermi level as the GNRs become wider, in our calculations it remains below the Fermi level without indications of a partial charge transfer to the substrate. In the absence of charge transfer, one would expect a rather symmetric closing of the gap. Thus, the observed behavior requires a width-dependent alignment of the aGNR levels with respect to those of Au(111). At this point it is worth noting that the dependence of the ribbon's polarizability on its width has been recently proposed as instrumental to understand the width dependence of the band gap of adsorbed GNRs.[29] However, in contrast to our experimental observations, this effect should affect both the VB and CB in a comparable way.

Figures S4 and S5 show the relaxed geometry and the distribution of the induced charge upon adsorption of the 6-aGNR on Au(111), respectively. The computed equilibrium adsorption height of the central portion of the ribbon is in the range of 3.23-3.34 Å (larger for wider ribbons). The limit for extended graphene is 3.34 Å, in good agreement with previous calculations.[43] Figure S4 shows how the electron charge accumulated in the surface as a result of the "pillow effect" extends beyond the region covered by carbon, piling up along the edges. This is most probably favored by the positive partial charge on the hydrogen atoms along the GNR edges. Thus, we find that the charge distribution is noticeably affected by the finite width of the ribbons. This characteristic distribution of the induced charge produces a width-dependent decrease of the interface dipole Δ, favoring a stronger p-type alignment the narrower is the GNR. This effect partially compensates the downshift of the VB as the GNRs become narrower, while increases the upshift of the CB. Therefore it can explain the asymmetric level movement observed in the DFT calculations and, as a consequence, be a key ingredient to understand also our experimental results. However, this effect smoothly scales inversely proportional to the GNR width, while the experiment evidences a relatively sharp transition from a rapidly varying valence band alignment at widths below 6-aGNRs to an almost unchanged valence band as the width increases (see Figure 3). In fact, as will be shown below, the transition seems to relate directly to the GNR´s band gap, which is no longer inversely proportional to the width if different families of armchair GNRs or even differently oriented GNRs are considered. Thus, in addition to the width-dependent interface dipole, additional bandgap-dependent effects must be responsible for our observations.

A sharp transition from having a varying onset energy in frontier band´s to having them quasi unchanged (close to the Fermi energy) is typically termed as Fermi level pinning. It is a well-known phenomenon observed at metal-semiconductor interfaces when one of the semiconductor´s bands gets close in energy to the substrate´s Fermi level, typically due to particularly high or low work functions (Fig. S6a). Observing a transition from a non-Fermi level pinning behavior to the pinning usually requires to change the work function and thereby the interface chemistry. Instead, in this particular study on Fermi level pinning the same phenomenology is observed while keeping the interface chemistry almost unchanged, the only varying parameter being the GNR´s width and thus its associated band gap (Fig. S6b).

The typically proposed explanation for Fermi level pinning is as follows. In the current case of an unchanged substrate and a decreasing adsorbate´s band gap, valence and conduction band onsets plotted vs. the band gap would be expected to symmetrically approach the mid-gap value with a slope S= 0.5 (Fig. 5a). However, as one of the bands (*e.g.* valence band in Fig. 5a) get close to the Fermi level, further band gap reductions create a compensating dipole moment σ to refrain the band from crossing the Fermi level. Although not confirmed by our calculations, whether due to the inability of current semilocal functionals to describe the energies of the electronic levels with sufficient accuracy for the free-standing ribbons and adequately incorporate correlation effects upon adsorption, or to slight deviations in the calculated adsorption distances that can substantially alter the interface energetics,[43] this is typically assumed to occur through partial charge transfer into metal-induced gap states.[47–49] As a result, the non-pinned band (conduction band in Fig. 5a) now supports a shift equal to the full band gap decrease, while the alignment of the pinned band remains unchanged (S=0). Avoiding the semiconductor´s bands to cross the Fermi level, the interfacial charge transfer is reduced. It is therefore commonly observed with physisorbed materials, since its absence would imply substantial charge transfers and consequently a chemisorption scenario.[50] This is exactly what we could expect from the interaction of graphene nanoribbons with an inert surface like Au(111).

Indeed, we observe a striking agreement between the model Fermi level pinning behavior of band onset vs. band gap (Fig. 5a) and the experimentally observed values from this work, together with results from other Au(111)-supported GNRs reported elsewhere, like 13-aGNR,[11] 9-aGNRS,[12] 7-aGNR,[33,34] 5-aGNRs,[51] and even (3,1) chiral GNRs [52] (Fig. 5b). As displayed in Fig. 4b, for band gap values above the critical value of ~1.7 eV, valence and conduction band shift symmetrically around their mid-gap value with a slope close to S=0.5. However, as the band gap gets below that threshold value, the valence band remains almost constant (S=-0.08±0.05) while the conduction band supports almost the full band decrease by shifting with a slope of S=0.92±0.05. Several GNRs have been characterized previously on an even less interacting substrate, namely on Au(111) with an intercalated Si layer. On that substrate the GNRs display a slightly larger bandgap due to the reduced substrate screening. However, not only does the band alignment similarly fit into our proposed Fermi level pinning picture, as expected from such weakly interacting surface, but it actually displays strikingly similar energies [30] (Fig. S7). Fermi level pinning energies of molecular adsorbate´s orbitals may vary substantially from system to system, which is normally associated with the different density of states hosted by the frontier orbitals and how far in energy the tails of the frontier bands´ density of states extend into the gap. In general, typical values remain below 0.4 eV away from the Fermi level,[53,54] which coincides with our observations.

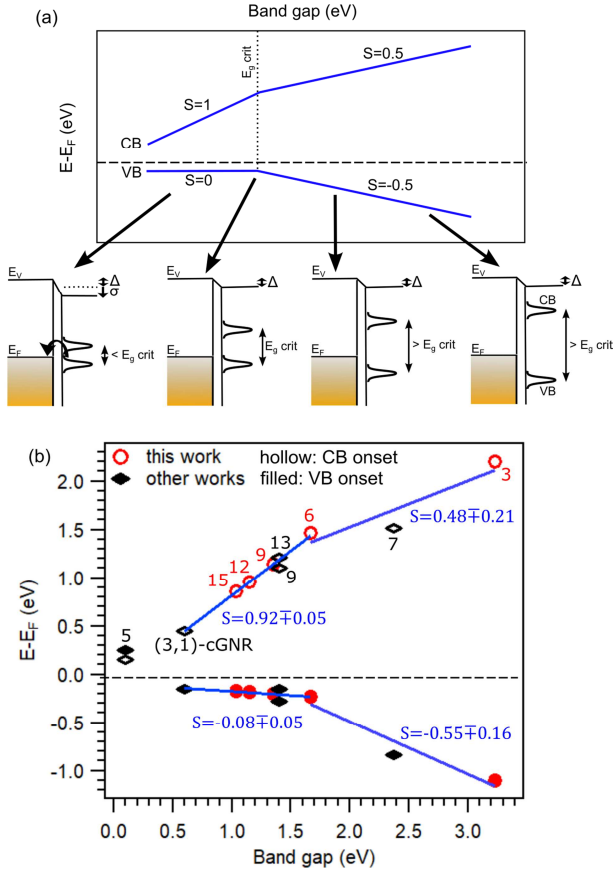

**Figure 5**. **(a)** Schematic evolution of the energy level alignment of valence and conduction band for varying adsorbate band gap. The energy level diagrams depict the interface dipole Δ from the molecular adsorption, as well as the additional dipole σ responsible for Fermi level pinning as the band gap gets narrower and one of the bands approaches the Fermi level (the valence band in this figure). It also displays the changing slopes of valence and conduction band onset vs. band gap as Fermi level pinning sets in. **(b)** Valence and conduction band onsets of GNRs studied in this work and in other reports of Au(111)-supported ribbons vs. their respective band gap (width given by numbers next to the CB symbols). Linear fits in selected regions display the changes in slope, evidencing notable similarity with the model scenario of panel (a).

Although still maintaining the same trend of a reduced slope of the valence band shift and an increased slope for the conduction band shift, 5-aGNRs deviate somewhat from the tendency observed for the rest of GNRs and even show a valence band onset slightly above the Fermi level. This is due to the small band gap of GNRs belonging to the $3p+2$ family.[51] They have correspondingly minute effective masses on their frontier bands [55] and thus a very low density of states around the gap. As a consequence, in a similar way as observed for graphene on a variety of substrates due to its associated low density of states around the Dirac point,[38,43] the bands of 5-aGNRs can cross the Fermi level with very little associated charge transfer and thus without creating substantial interface dipoles. Instead, the case of nanoribbons with sizeable band gaps is utterly different in that it exhibits a large density of states at the van Hove singularity of the valence band onset, causing a clearer Fermi level pinning.

**CONCLUSIONS**

In conclusion, we report spectroscopic evidence of the width-dependent band gap predicted for armchair graphene nanoribbons a decade ago and the associated energy level alignment. Starting with the synthesis of PPP wires (3-aGNRs), subsequent annealing drives their lateral fusion into 6-, 9-, 12-, or 15-aGNRs depending on the number of involved PPP chains. That is, the first 5 members of *3p*-aGNR family are obtained on the same surface, on which both valence and conduction band are probed by means of STS. We observe a continuously decreasing band gap as the GNRs structures get wider. Most importantly, Fermi level pinning of the valence band is found on Au(111) for 6 or more dimer lines wide aGNRs, in qualitative agreement with DFT calculations. Results known from other GNRs equally fit the pinning phenomenon proposed here, whenever their bandgap is below ~1.7 eV. This has important implications on the energy level alignment across GNR/metal interfaces, which may in turn be crucial for future GNR-based devices displaying similar interfaces at charge collection electrodes.

**METHODS**

Experimental Procedures

Samples were prepared by deposition of precursor molecules from a Dodecon molecular evaporator containing a Knudsen cell where the molecular precursors are placed and heated up to a sublimation temperature of 400 K. We used a single monocrystalline Au(111) surface as substrate. Standard sputtering/annealing cycles (1.5kV Ne, 730K) were performed to obtain an atomically clean Au (111) surface. We grew PPP nanowires and aGNRs by initial DBTP deposition and subsequent annealing at 250°C and 430°C, respectively, in the $10^{-10}$ mbar range. The annealing processes were performed in such a way that the maximum temperature of 250 ºC or 430 ºC was maintained for 30 minutes, respectively. STM measurements were performed using a home-built STM with the samples held at 4.2 K in UHV conditions. A tungsten tip was used for topography and spectroscopic measurements. Topography was obtained in constant current mode of the STM. For spectroscopic measurements, the tunneling differential conductance was measured by a lock-in amplifier, while the sample bias was modulated by a 767 Hz, 12-18 mV (rms) sinusoidal voltage under open-feedback conditions.
All STM images were processed by WSxM.[56]

Computational Procedures

The electronic structure and geometries were calculated using density functional theory, as implemented in the SIESTA code.[57] We use a supercell description of the system, made up of a slab containing 4 layers of Au(111), with a graphene nanoribbon of 3, 6, 9 or 12 dimer lines on the top Au surface. We employed the stacking geometry suggested in ref. 43, where 2x2 graphene and Au(111) √3×√3 unit cells are directly matched (more details of the set up in the Supp. Inf.). The bottom Au surface was passivated with hydrogen atoms to quench one of the Au(111)'s Shockley surface states close to the Fermi level.[58] In order to properly capture the "pillow effect", it is instrumental to describe adequately the decay of the metal electron density into vacuum and the subsequent modifications of the surface dipole layer upon adsorption of the ribbon. The use of a basis set of extended, optimized atomic orbitals to describe the Au surface atoms is thus crucial to capture the correct level alignment. To avoid interactions between periodic surfaces, we considered a vacuum region of more than 10 Å. The GNRs and the top two Au layers were fully relaxed until forces were < 0.01eV/Å, and the dispersion interactions were taken into account by the non-local optB88-vdW functional.[59] The basis set consisted of double-zeta plus polarization (DZP) orbitals for C, H and bulk Au atoms. Au atoms on the topmost layer were treated with a DZP basis-set optimized for the description of the (111) surface of Au.[60] A 9x1x1 Monkhorst-Pack mesh was used for the k-point sampling of the three-dimensional Brillouin zone, where the 9 k-points are taken along the direction of the ribbon. A cutoff of 300 Ry was used for the real-space grid integrations.

ASSOCIATED CONTENT

**Supporting Information**. dI/dV point spectra on PPP nanowires and conductance maps at onset energies of valence and conduction band (Fig. S1). Conductance maps and associated profiles revealing the valence band of GNRs (Fig. S2). Point spectra and conductance maps in a wider energy range on aGNRs of varying width, displaying the onset of the second valence band (VB-1) (Figure S3). Example of Au(111)-GNR unit cell used for DFT calculations (Figure S4). Iso-surface of the computed induced charge upon adsorption of a 6-aGNR on Au(111) (Figure S5). Schematic model of the changing energy level alignment of weakly interacting adsorbate´s valence and conduction bands with varying work function of the substrate for fixed adsorbate´s band gap, compared to the alignment with a fixed work function and a varying adsorbate´s band gap (Figure S6). Experimental data of valence and conduction band onset of GNRs studied in this work, as well as for works reported elsewhere on other Au(111)-supported ribbons and ribbons on Au(111) with an intercalated Si layer.

This material is available free of charge *via* the Internet at http://pubs.acs.org.

AUTHOR INFORMATION

**Corresponding Author**

d_g_oteyza@ehu.es

**Author Contributions**

The manuscript was written through contributions of all authors. All authors have given approval to the final version of the manuscript.

ACKNOWLEDGMENT.

The project leading to this publication has received funding from the European Research Council (ERC) under the European Union's Horizon 2020 research and innovation programme (grant agreement No 635919), from the European Union FP7 FET-ICT ``Planar Atomic and Molecular Scale devices'' (PAMS) project (Contract No. 610446), from the Spanish Ministry of Economy,

Industry and Competitiveness (MINECO, Grant No. MAT2016-78293-C6) and from the

University of Padova (Grant CPDA154322, Project AMNES).

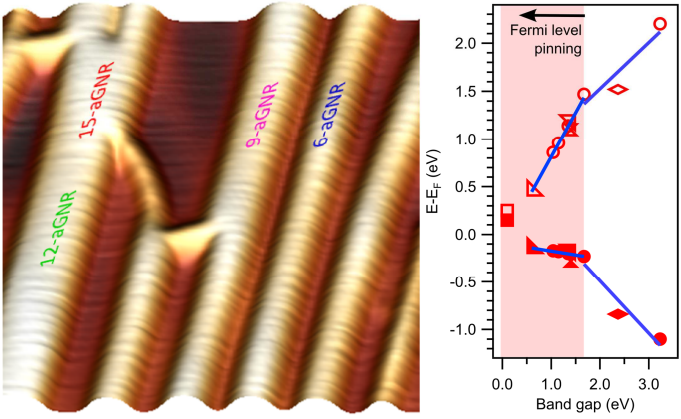

Insert Table of Contents Graphic and Synopsis Here

# Supplementary Information:

# Width-Dependent Band Gap in Armchair Graphene Nanoribbons Reveals Fermi Level Pinning on Au(111).

*Néstor Merino-Díez,* [‡,†] *Aran Garcia-Lekue,* [‡,§] *, Eduard Carbonell-Sanromà,* [†] *Jincheng Li,* [⊥,†] *Martina Corso,* [†,§,⊥] *Luciano Colazzo,* [∇] *Francesco Sedona,* [∇] *Daniel Sánchez-Portal,* [‡,⊥] *Jose I. Pascual,* [†,§] *and Dimas G. de Oteyza* [‡,§,⊥].

[‡] Donostia International Physics Center (DIPC), 20018 Donostia – San Sebastián, Spain

[†] CIC nanoGUNE, Nanoscience Cooperative Research Center, 20018 Donostia – San Sebastián, Spain

[§] Ikerbasque, Basque Foundation for Science, 48013 Bilbao, Spain

[⊥] Centro de Física de Materiales (CSIC/UPV-EHU) - Materials Physics Center, 20018 Donostia – San Sebastián, Spain

[∇] Dipartimento di Scienze Chimiche, Università di Padova, 35131 Padova, Italy

Author(s) address: d_g_oteyza@ehu.es

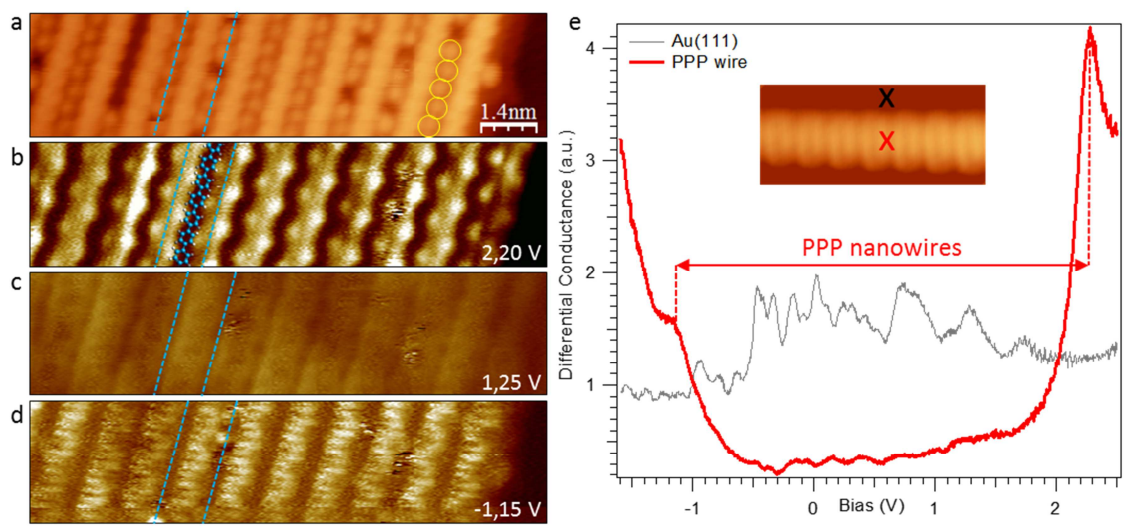

**Figure S1. (a)** STM topography image (12.3 nm x 2.9nm; $V_s = -1.1$ V; $I_t = 0.60$ nA). Conductance maps (12.3 nm x 2.9nm; $I_t = 0.60$ nA) on PPP wires at **(b)** valence band edge ($V_s = 2.20$ V), where simulated representation of PPP wire is included as visual guide; **(c)** within the band gap ($V_s = 1.25$ V), and **(d)** at conduction band edge ($V_s = -1.15$ V). Blue lines superimposed to the adsorbated Br atoms (yellow circles) to the substrate are included for an easier appreciation of single nanowire features. **(e)** Spectra taken on PPP nanowires (in red) where Au(111) signal (in black) is added to every spectrum as background reference (open-feed-back parameters: $V_s = -1.60$ V; $I_t = 0.54$ nA; modulation voltage $V_{rms} = 18$ V). Crosses in inset topographic image indicate the positions where spectra were recorded.

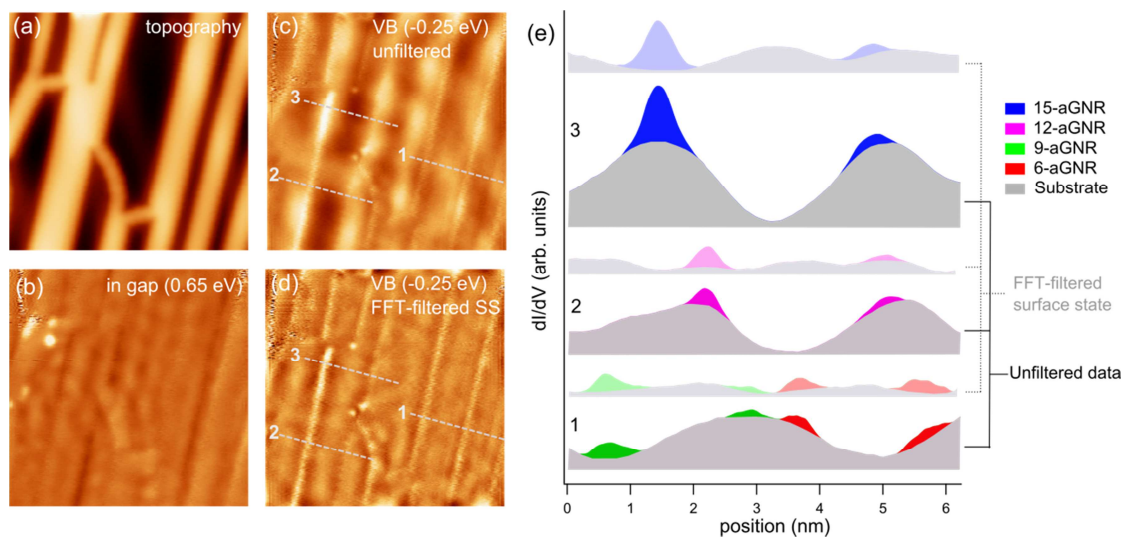

**Figure S2.** (a) STM topography image (12,4 nm x 12.4 nm; $V_s = -1.1$ V; $I_t = 0.61$ nA) displaying GNRs of varying width. (b) Conductance map at an energy within the band gap ($V_s = 0.65$ V), revealing the absence of GNR-related signal and only the oscillations from the scattered Au surface state. (c) Conductance map near the valence band onset ($V_s = -0.25$ V). The three numbered lines mark the position of profiles taken so as to cross GNRs of the different studied widths. (d) Same conductance map as in (c) but after filtering the main surface state signal. To do so, an FFT image of (c) is calculated, the k corresponding to the surface state is cut and then the image is transformed back. Without most of the substrate-related signal, the overall corrugation is lowered, making it easier to distinguish the GNR-related signal (e) Profiles across the conductance maps in (c) and (d), highlighting the contribution from the VB of differently wide GNRs on top of the underlying substate contribution. Comparing the profiles on filtered and unfiltered conductance images one can clearly distinguish how by applying the filter the GNR signal remains unaltered on top of a substantially less corrugated substrate signal, overall making its distinction easier.

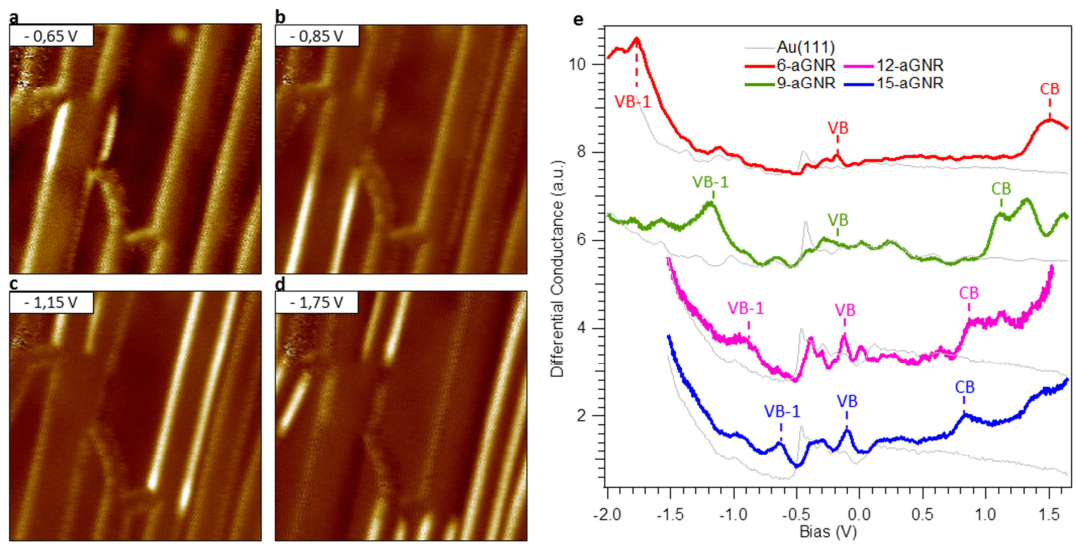

**Figure S3. (a-d)** Conductance maps (12.4 nm x 12.4 nm; $I_t$ = 0.60 nA) at the onsets of the second valence bands (VB-1) on **(a)** 15-aGNR ($V_s$ = -0.65 V), **(b)** 12-aGNR ($V_s$ = -0.85 V), **(c)** 9-aGNR ($V_s$ = -1.15 V) and **(d)** 6-aGNR ($V_s$ = -1.75 V). **(e)** Spectra taken on 6-aGNR (in red), 9-aGNR (in green), 12-aGNR (in pink) and 15-aGNR (in blue) where Au(111) signal (in black) is added to every spectrum as background reference (open-feed-back parameters: $V_s$ = -1.0 V; $I_t$ = 0.5–10 nA; modulation voltage $V_{rms}$ = 10-12 mV).

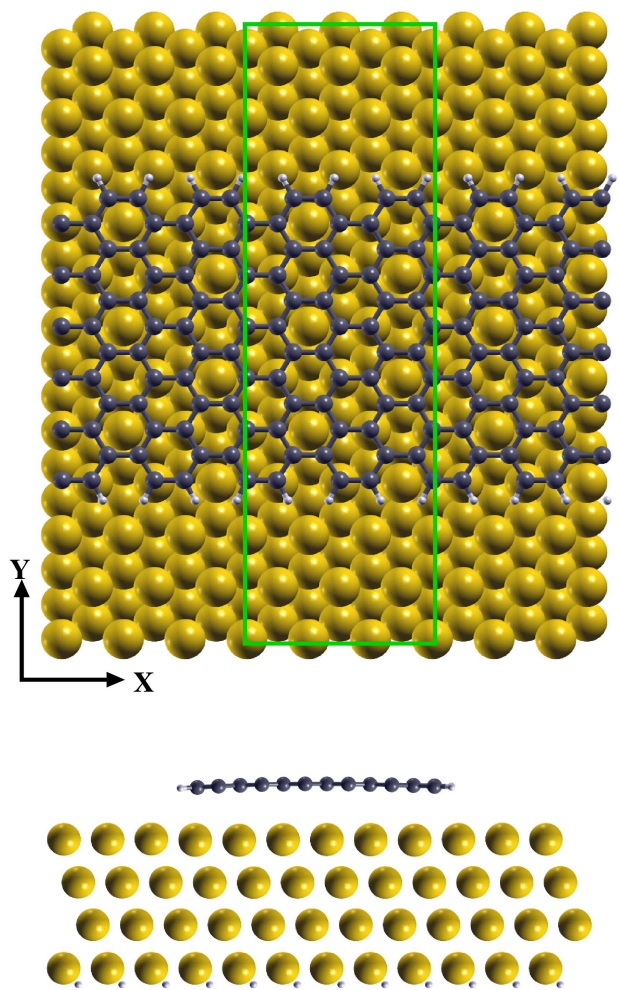

**Figure S4.** Example of one of the setups used for the calculations: graphene nanoribbon of 12 dimer lines on a Au(111) slab with a lateral unit cell as indicated by the green rectangle. The ribbon is periodic along the X direction.

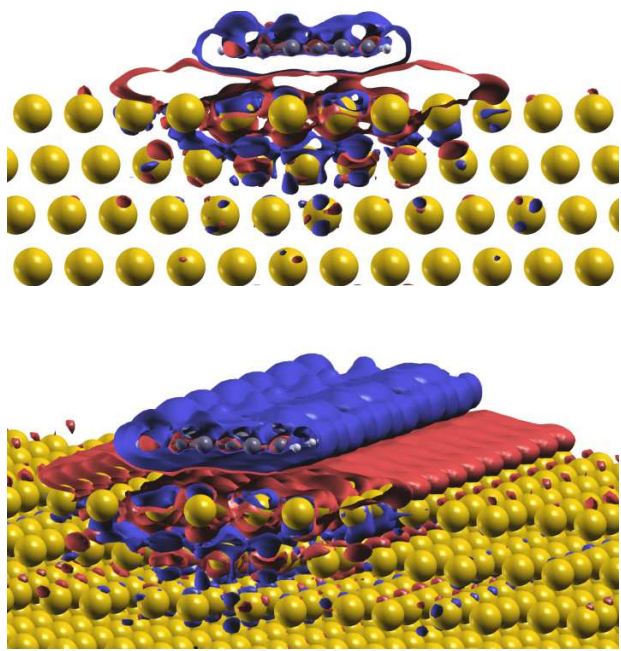

**Figure S5.** Iso-surface ($5\times10^{-4}$ electrons/bohr$^3$) of the computed induced charge upon adsorption of a 6-aGNR on Au(111). **Top:** front view. **Bottom:** side view. Red surfaces correspond to electron accumulation, while blue surfaces correspond to electron depletion. Pauli repulsion by the ribbon electrons pushes the electrons of gold towards the surface, giving rise to a modification of the surface dipole. As clearly shown in the image, the electron accumulation spills out from the region covered by the ribbon, modifying the surface dipole with respect to that of extended graphene. The modification will be larger the narrower the ribbon.

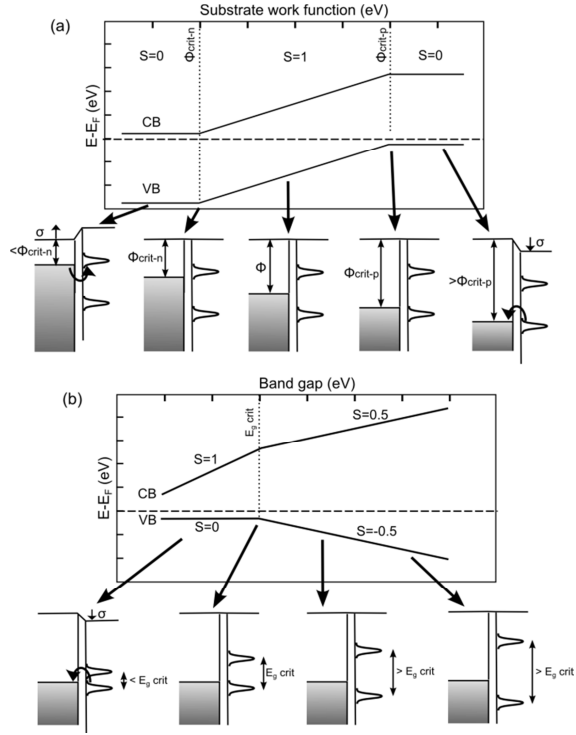

**Figure S6**. **(a)** Schematic model of the changing energy level alignment of weakly interacting adsorbate´s valence and conduction bands with varying work function of the substrate. The alignment changes from vacuum level pinning to Fermi level pinning of valence and conduction bands at critically high and low work functions, respectively. Analyzing the slope S of the energy level alignment changes vs. the work function, this translates in changes from S=1 for vacuum level pinning to S=0 for Fermi level pinning. **(b)** Schematic model of the changing energy level alignment of weakly interacting adsorbate´s valence and conduction bands with varying band gap. As the band gap is reduced, the valence and conduction band onsets move symmetrically towards the mid-gap value with a slope S=0.5. As one of the bands gets sufficiently close to Fermi, Fermi level pinning of that band sets in (valence band in the picture, S=0) and only the non-pinned band supports the full band gap change henceforth, shifting accordingly with a slope of S=1.

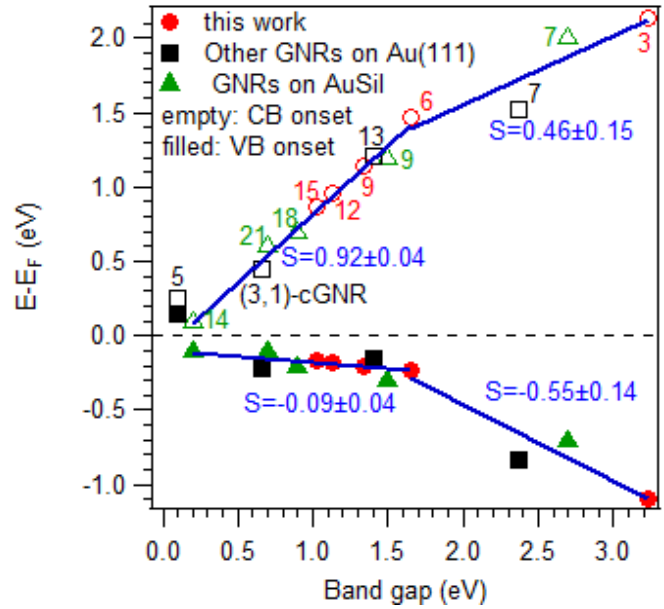

**Figure S7**. Valence and conduction band onsets of GNRs studied in this work and in other reports of Au(111)-supported ribbons vs. their respective band gap (width given by the numbers next to CB symbols). Data of GNRs on an even less interactive substrate as is a Si intercalated layer on Au(111) are also added for comparison. Linear fits in selected regions display the changes in slope, evidencing remarkable similarity with the model scenario of Fig. S6b. The data not corresponding to this work have been taken from references [1] (7-aGNR/Au(111)), [2] (9-aGNR/Au(111)), [3] ((3,1)-cGNR/Au(111)), [4] (5-aGNR/Au(111)) and [5] (aGNRs/AuSil).